\documentclass[a4paper,fleqn,usenatbib]{mnras}
\usepackage{graphicx}	
\usepackage{amsmath}	
\usepackage{amssymb}	
\usepackage{multicol}        
\usepackage{bm}		
\usepackage{pdflscape}	

\usepackage{multirow}
\usepackage{textcomp}
\usepackage{amssymb}

\usepackage[T1]{fontenc}
\usepackage{ae,aecompl}

\usepackage[utf8]{inputenc} 

\newcommand{\cxo}{{\em Chandra}}

\newcommand{\xmm}{{\em XMM--Newton}}

\def\nh {$N_{H}$}
\def\chisq {$\chi ^{2}$}
\def\rchisq {$\chi_{\nu} ^{2}$}
\def\ergs {erg\,s$^{-1}$}
\def\ergscm2 {erg\,s$^{-1}$cm$^{-2}$}
\def\ss {s\,s$^{-1}$}
\def\cm2 {cm$^{-2}$}
\def\arcsec{$^{\prime\prime}$}

\def\kt{$kT_{\rm BB}$}
\def\r{$R_{\rm BB}$}
\def\xdin{RX\,J1308.6$+$2127}

\title[Narrow phase-dependent features in XDINSs]{Narrow phase-dependent features in X-ray Dim Isolated Neutron Stars: a new detection and upper limits}

\author[A. Borghese et al.]{
A. Borghese$^{1}$\thanks{E-mail: a.borghese@uva.nl},
N. Rea$^{1,2}$, F. Coti Zelati$^{1,3,4}$, A. Tiengo$^{5,6,7}$, R. Turolla$^{8,9}$, S. Zane$^{9}$\\
$^{1}$Anton Pannekoek Institute for Astronomy, University of Amsterdam, Postbus 94249, NL--1090 GE Amsterdam, The Netherlands\\
$^{2}$Institute of Space Sciences (IEEC--CSIC), Campus UAB, Carrer Can Magrans s/n, 08193 Barcelona, Spain\\
$^{3}$Universit\`a dell'Insubria, via Valleggio 11, I--22100 Como, Italy\\
$^{4}$INAF--Osservatorio Astronomico di Brera, via Bianchi 46, I--23807 Merate (LC), Italy\\
$^{5}$Scuola Universitaria Superiore IUSS Pavia, piazza della Vittoria 15, I--27100 Pavia, Italy\\
$^{6}$Istituto Nazionale di Fisica Nucleare, Sezione di Pavia, via A. Bassi 6, I--27100 Pavia, Italy\\
$^{7}$INAF--Istituto di Astrofisica Spaziale e Fisica Cosmica, via E. Bassini 15, I--20133 Milano, Italy\\
$^{8}$Dipartimento di Fisica e Astronomia, Universit\`a di Padova, via F. Marzolo 8, I--35131 Padova, Italy\\
$^{9}$Mullard Space Science Laboratory, University College London, Holmbury St. Mary, Dorking, Surrey RH5 6NT, UK}

\date{Accepted XXX. Received YYY; in original form ZZZ}

\pubyear{2016}

\begin{document}
\label{firstpage}
\pagerange{\pageref{firstpage}--\pageref{lastpage}}
\maketitle

\begin{abstract}

We report on the results of a detailed phase-resolved spectroscopy of archival \xmm\, observations of X-ray Dim Isolated Neutron Stars (XDINSs). Our analysis revealed a narrow and phase-variable absorption feature in the X-ray spectrum of \xdin . The feature has an energy of $\sim$740\,eV and an equivalent width of $\sim$15\,eV. It is detected only in $\sim$ 1/5 of the phase cycle, and appears to be present for the entire timespan covered by the observations (2001 December - 2007 June). The strong dependence on the pulsar rotation and the narrow width suggest that the feature is likely due to resonant cyclotron absorption/scattering in a confined high-B structure close to the stellar surface. Assuming a proton cyclotron line, the magnetic field strength in the loop is B$_{loop} \sim 1.7 \times 10^{14}$\,G, about a factor of $\sim$5 higher than the surface dipolar magnetic field (B$_{surf} \sim 3.4 \times 10^{13}$\,G). This feature is similar to that recently detected in another XDINS, RX\,J0720.4-3125, showing (as expected by theoretical simulations) that small scale magnetic loops close to the surface might be common to many highly magnetic neutron stars (although difficult to detect with current X-ray instruments). Furthermore, we investigated the available \xmm\, data of all XDINSs in search for similar narrow phase-dependent features, but could derive only upper limits for all the other sources.

\end{abstract}


\section{Introduction}

Thanks to its high sensitivity in the soft X-ray band (0.1--2.5 keV), \textit{ROSAT} led to the discovery of seven thermally emitting X-ray pulsars, known as X-ray Dim Isolated Neutron Stars (XDINSs; see \citealt{2009ASSL..357..141T} for a review). Timing studies have shown that all the XDINSs rotate slower ($P\sim 3$--$11$\,s) and have higher inferred surface dipolar magnetic fields ($B_{\rm dip}\approx 10^{13}$\,G) than the bulk of the radio pulsars (\citealt{2011ApJ...740L..30K}; and references therein); radio emission has not been detected so far despite deep searches \citep{2009ApJ...692K..702}, but all of them have confirmed optical and ultraviolet counterparts \citep{2011ApJ...736..117K}. They are considered to be steady sources, apart from RX\,J0720.4-3125 which is the only one that exhibits long-term variations in its timing and spectral properties \cite[][and references therein]{2012MNRAS.423.1194H}. The low values of the column density (\nh\ $\approx 10^{20}$\,cm$^{-2}$) derived from X-ray data make XDINSs among the closest known neutron stars, with distance of a few hundred parsecs ($\lesssim$ 500\,pc, see \citealt{2007Ap&SS.308..171P}). These values are compatible with the distances inferred from the parallax measurements, available up to now only for two XDINSs \citep{2007Ap&SS.308..191V}. 
Their X-ray luminosity $L_{\rm X}\approx 10^{31-32}$\,\ergs\, exceeds the spin-down luminosity, making XDINSs different from the rotation-powered pulsars. They have estimated ages of a few 10$^5$\,yr, derived from kinematics and cooling curves with magnetic field decay, while their characteristic age is longer ($\approx 10^6$\,yr).\\


Their X-ray spectra show only thermal emission without the hard power-law component often observed in other isolated neutron stars. This thermal emission is thought to be due to residual heat and to come directly from the stellar surface with neither significant evidence for magnetospheric activities nor contamination by rotationally-powered non-thermal contribution. For these reasons XDINSs are the ideal targets for probing neutron star surface emission properties. However, comparing theoretical predictions with observations is challenging, despite efforts in this direction have been carried out by several authors \cite[e.g.][]{2007MNRAS.375..821H,2008ApJS..178..102H}. At present, a self-consistent model of the multiwavelength spectral energy distribution of XDINSs is still lacking and the surface composition, the thermal and magnetic map of these stars remain unknown.
  
An absorbed blackbody model provides a good description for their spectral energy distribution in the soft X-ray band with inferred temperature in the range $kT \sim$ 50--100\,eV. However, in the last decade \xmm\ and {\it Chandra} observations have detected deviations from a pure Planckian distribution in all of the XDINSs, except for RX\,J1856.5-3754\footnote{For RX J0420.0-5022 an absorption line at $\sim 0.3$ keV was reported by \cite{2004A&A...424..635H}, but not confirmed by later observations \citep{2011ApJ...740L..30K}.}. These spectral features are typically modelled by a Gaussian profile in absorption centred at energies of some hundreds eVs; the width of the lines is $\sim$70--170\,eV and the equivalent width ranges between $\sim$ 30 and 150\,eV. The origin of these broad absorption features is still under debate: they can be produced by proton cyclotron resonances/atomic transitions in a magnetized atmosphere \citep{2007Ap&SS.308..191V} or by an inhomogeneous surface temperature distribution \citep{2014MNRAS.443...31V}. A search for narrow absorption features in \xmm\, Reflection Grating Spectrometer (RGS) data was performed by \citet{2012MNRAS.419.1525H} for the four brightest XDINSs, identifying spectral lines at $\sim$ 0.56\,keV in three of them. 

Recently, a phase-dependent absorption feature, present only for $\sim$ 20\% of the pulsar rotation,  was detected in the X-ray spectrum of RX\,J0720.4-3125 \citep{2015ApJ...807L..20B}. Features with somewhat similar properties were previously reported in two low-field magnetars, SGR\,0418+5729 and SWIFT\,J1822.3-1606 \citep{2013Natur.500..312T, 2016MNRAS.456.4145R}, and, because of the strong dependence on the rotational phase, believed to be produced by proton cyclotron resonant scattering in a magnetic loop close to the star surface. According to magneto-thermal evolution models \citep{2013MNRAS.434..123V}, XDINSs are likely to be the descendants of magnetars, therefore we expect to find similar spectral features which would provide evidence for a surface magnetic field structure more complex than a pure dipole \citep{2005AdSpR..35.1162Z}. Moreover, XDINSs are nearby and very bright sources that have been monitored for a long timespan, so a large amount of data and detailed timing solutions are available for most of them.
These findings have motivated our search for such phase-dependent features in the X-ray spectra of all the other XDINSs.\\

\xdin , also known as RBS\,1223 and 1RXS\,J130848.6+212708, was identified as a possible nearby isolated neutron star by \citet{1999A&A...341L..51S} in the \textit{ROSAT} Bright Survey. Unlike other XDINSs, its rotational phase-folded light curves exhibit a remarkable double-humped shape. According to \citet{2005A&A...441..597S} the main source of this observed behaviour is an inhomogeneous temperature distribution aver the stellar surface. 
Moreover, \xdin\, has the largest pulsed fraction\footnote{The pulsed fraction is defined as PF $\equiv \frac{CR_{max} - CR_{min}}{CR_{max} + CR_{min}}$ where CR is the count rate.} among this class of INSs, $\sim$ 19$\%$ in the 0.2--1.2\,keV energy range; the PF increases with energy. No long-term change in the flux has been detected so far \citep{2005A&A...441..597S}.  


The X-ray phase-averaged spectrum is well fitted by a combination of an absorbed blackbody model with inferred temperature $\sim$ 85\,eV and a Gaussian absorption line at energy $\sim$ 270\,eV \citep{2011A&A...534A..74H, 2003A&A...403L..19H}. This absorption feature has the largest equivalent width among all XDINSs. \citet{2011A&A...534A..74H} performed a phase-resolved spectroscopy by fitting spectra corresponding to maxima and minima of the double-peaked light curve, and found a clear variation of the blackbody temperature and the line energy along the phase cycle, these being larger at the peaks. Alternatively, the phase-resolved spectra can be simultaneously fitted by a spectral model consisting of an iron condensed surface with a partially ionized hydrogen atmosphere on top. This fit provides constraints for the physical properties of the source, such as temperature and magnetic field strength at the poles, in addition to an estimate of the mass-to-radius ratio, $(M/M_\odot)(R/{\rm km})=0.087\pm0.004$, which suggests a stiff equation of state.\\

Here we re-analyse all the available \xmm\ observations of \xdin, performing a detailed phase-resolved spectral analysis. We report our results in Section \ref{sect:data}, and present a possible new phase-dependent absorption feature in its X-ray spectrum. In Section \ref{sect:MC}, we describe the Monte Carlo simulation procedure applied and how the result supports our discovery. Moreover we investigate \xmm\ data for the other XDINSs looking for similar features, but we could derive only upper limits, reported in Section \ref{sect:UL}. In Section \ref{sect:disc} we discuss the possible origin of the new feature observed in the spectrum of \xdin. Conclusions follow in Section \ref{sect:concl}.

\begin{table*}
\caption{Summary of the {\xmm}/EPIC-pn observations of RX J1308.6$+$2127.}
\label{tab:log}
\footnotesize{
\begin{tabular}{@{}lcccccc}
\hline
\hline
Obs. ID  		&  Obs. Date 		& Read-out mode$^a$ 	& Live time$^b$ 	 & Count rate$^c$ & Pile-up fraction ratio$^d$ & Pile-up fraction ratio$^d$		\\ 
			& YYYY-MM-DD 	& 					& (ks) 	& (counts s$^{-1}$)	& single pattern & double pattern \\
\hline
 0090010101 & 2001 Dec 31 & SW & 13.0 & 2.22(1) & 1.008(8) & 0.961(15) \\
 0157360101 & 2003 Jan 01 & FF & 24.2 & 2.25(1) & 0.982(6) & 1.124(13) \\
 0163560101 & 2003 Dec 30 & FF & 27.0 & 2.30(1) & 0.989(5) & 1.072(12) \\
 0305900201$^*$ & 2005 Jun 25 & FF & 13.2 &  2.35(1) & 0.986(8) & 1.095(17) \\
 0305900301$^*$ & 2005 Jun 27 & FF & 11.5 &  2.28(1)& 0.991(8) & 1.059(18) \\
 0305900401$^*$ & 2005 Jul 15 & FF & 11.4 &  2.25(1) & 0.986(8) & 1.093(18) \\
 0305900601 & 2006 Jan 10 & FF & 13.2 & 2.27(1) & 0.986(8) & 1.096(17) \\
 0402850301$^{**}$ & 2006 Jun 08 & LW & 4.8 & 2.22(2) & 0.991(13) & 1.063(27) \\
 0402850401$^{**}$ & 2006 Jun 16 & LW & 5.7 & 2.22(2) & 0.982(12) & 1.112(25) \\
 0402850501$^{**}$ & 2006 Jun 27 & LW & 9.4 & 2.26(2) & 0.991(9) & 1.055(19) \\
 0402850901$^{**}$ & 2006 Jul 05 & LW & 6.5 & 2.26(2) & 0.984(11) & 1.101(23) \\
 0402850701$^\circ$ & 2006 Dec 27 & LW & 7.6 & 2.23(2) & 0.988(10) & 1.081(22) \\
 0402851001$^\circ$ & 2007 Jun 11 & LW & 8.0 & 2.35(2) & 0.990(1) & 1.064(20) \\
 	
\hline
\hline
\end{tabular}}

\begin{list}{}{}
\item[$^a$] FF: full-frame (time resolution of 73 ms); SW: small window (time resolution of 6 ms); LW: large window (time resolution 48 ms). 
\item[$^b$] Live time refers to the duration of the observations after correcting for instrumental dead-time and filtering for background flares.
\item[$^c$] Count rates refer to the spectra extracted with a circle region with PATTERN $=$ 0 in the 0.1--10 keV energy band; errors are quoted at the 1$\sigma$ confidence level.
\item[$^d$] Observed-to-model singles and doubles pattern pile-up fraction ratios are calculated in the 0.1--2 keV energy range for a circular extraction region of radius 30\arcsec\, using the \texttt{SAS epatplot}. Errors are quoted at the 1$\sigma$ confidence level. 
\item[$^{*,**}$] The labelled observations were merged in the spectral analysis. 
\item[$^\circ$] These observations were not considered in the spectral analysis, since they had insufficient exposure time. 
\end{list}
\end{table*}

\section{XMM-Newton analysis}
\label{sect:data}

\subsection{Observations and data reduction}
\label{subsect:reduction}

Thirteen observations of \xdin\, were carried out by the \xmm\, satellite using the European Photon Imaging Camera (EPIC) between 2001 December 31 and 2007 June 11. Although different operating modes were adopted, all the available observations were performed with the thin optical blocking filter; a log of them is given in Table \ref{tab:log}. We re-analysed here only the data acquired with the EPIC-pn camera \citep{2001A&A...365L..18S} because they are less affected by pile-up and provide spectra with higher counting statistics than MOS data (the estimated pile-up level for single pixel events for MOS data is always greater than 10\%). We processed the raw data using the \texttt{epproc} task, following the standard threads from the Science Analysis Software (\texttt{sas}, version 15.0.0) with the most up to date calibration files available. We removed any particle flares via good-time-intervals. We adopted the coordinates reported by \citet{2002ApJ...579L..29K} for the X-ray position, i.e. $\rm RA=13^h08^m48\fs27$, $\rm Dec= +21^\circ27'06\farcs8$ (J2000.0), to convert the photon arrival times to Solar System barycentre reference frame, by means of the \texttt{barycen} task. A rotational phase was assigned to the source counts of all the observations using the timing solution provided by \citet{2005ApJ...635L..65K}; the 10.31\,s pulsations are slowing down at a rate of $\dot{P}$ = 1.120(3) $\times$ 10$^{-13}$\,\ss . In the following uncertainties are quoted at the 90 per cent confidence level for a single parameter of interest ($\Delta$\chisq  = 2.706), unless otherwise noted.

For all the observations, we extracted the source photons from a circular region of radius 30\arcsec\, centered on the source point spread function (PSF) and the background counts through the same circle far from the source location, but on the same CCD. We checked for the potential impact of the pile-up level through the \texttt{epatplot} tool; in order to mitigate it, we restricted our spectral analysis to photons having FLAG=0 and PATTERN=0. We generated the redistribution matrices with the \texttt{rmfgen} tool, including an additional correction to further minimise the spectral distortions caused by pile-up\footnote{http://www.cosmos.esa.int/web/xmm-newton/sas-thread-epatplot}, and ancillary response files with the \texttt{arfgen} tools for each spectrum.
Before fitting, the background-subtracted spectra were rebinned according to a minimum number of 100 counts per spectral bin and not to oversample the spectral energy resolution by more than a factor of 5; this was performed with the SAS tool {\tt specgroup} fixing the parameter {\tt oversample} equal to 5. The spectral modelling was performed within the XSPEC analysis package (version 12.9.0, \citealt{1996ASPC..101...17A}), using the \chisq\, statistics.

To improve the statistics we decided to merge the observations performed during 2005, these being only 20 days apart, and obtained a spectrum with a total exposure time of 35.1\,ks. Moreover, we built a combined spectrum with a total exposure time of 25.0\,ks, comprising the observations carried out during June--July 2006. We used the \texttt{merge} tool to merge the corresponding EPIC event lists, taking care of combining only event files with the same instrumental set-up, to minimise systematic errors. Spectra relative to observations with an exposure time less than 10\,ks were excluded from our analysis because of insufficient statistics. We hence focused on six spectra in total.

\subsection{Phase-averaged spectral analysis}

To fit the phase-averaged spectra we used a combination of an absorbed blackbody model (\texttt{bbodyrad} in XSPEC) and a Gaussian absorption line (additive model \texttt{gauss}). To describe the absorption by the interstellar medium along the line of sight, the \texttt{phabs} model was adopted with cross-sections from \citet{1992ApJ...400..699B} and solar chemical abundances from \citet{1989GeCoA..53..197A}. We note that the choice of the abundance and cross-sections tables does not affect the results of the spectral fitting (all parameters are compatible within 90\% confidence level), due to the low photoelectric absorption in the source direction. We fitted the six phase-averaged spectra in the energy range 0.2--1.2\,keV, first individually. Additionally, to better constrain the spectral parameters and reduce the number of degrees of freedom (dof hereafter), we performed a simultaneous fit of the six spectra where the relative normalizations between different operating modes were allowed to vary and a systematic error of 1.5\% ( {\tt syst} parameter in XSPEC) was assigned to each spectral bin.  This is an energy-independent systematic error added to the model in XSPEC and accounts for cross-calibration uncertainties\footnote{http://xmm2.esac.esa.int/docs/documents/CAL-TN-0018.pdf}. We preferred not to merge all the observations to create a combined spectrum, because different observational modes were used for the EPIC-pn camera, and this procedure introduces additional systematic errors\footnote{ In our previous work on RX J0720.4-3125 \citep{2015ApJ...807L..20B}, we merged all the spectra, because the observations were performed with the same modes.}.




The best fit yielded the following parameters: hydrogen column density \nh\ $= (3.0\pm1.2) \times 10^{20}$\cm2 , blackbody temperature \kt\ = 83.9$\pm$1.1\,eV, averaged blackbody radius \r\ = 3.2(2)\,km, line energy E$_1$ = 203$_{-37}^{+40}$\,eV, width $\sigma_1$ = 139$_{-13}^{+12}$\,eV and normalization of -2.9$^{+0.9}_{-0.8} \times 10^{-2}$. From hydrogen column density, the distance of \xdin\, is estimated to be at least 180 pc \citep{2010MNRAS.402.2369T}. We assumed this lower value to estimate the blackbody radius. The \nh\ value obtained is compatible with the total Galatic absorption in the source direction (\nh\ $= 2.4 \times 10^{20}$\cm2 ; \citealt{2013MNRAS.431..394W}) and with the value estimated by \citet{2003A&A...403L..19H}. 

The equivalent width of the broad feature is 179$^{+3}_{-59}$\,eV, confirming it is the highest among the XDINSs. The above-mentioned fit leads to a reduced chi-square \rchisq = 1.39 for 180 dof. The absorbed and unabsorbed flux in the 0.1--10\,keV band are 3.39$_{-0.02}^{+0.04} \times 10^{-12}$ \ergscm2 and 7.32$_{-0.41}^{+0.21}\times 10^{-12}$ \ergscm2 , respectively, estimated with the {\tt cflux} model in XSPEC.

Note that the energy we found for the broad feature is different from what was derived in previous works. Hambaryan et al. (2011) built a combined spectrum, merging all the observations except that performed in Small Window mode, and used the multiplicative component model \texttt{gabs} to fit the broad feature, obtaining a higher value for the line energy. These different choices in the analysis (possibly also the different assumptions in the abundances and cross section of the photoelectric absorption model) might be the source of the differences in the broad line properties.




\subsection{Phase-resolved spectroscopy and dynamic phase-energy images}
\label{subsect:pha_resol}

\begin{figure*}
\begin{center}
\includegraphics[scale=0.5]{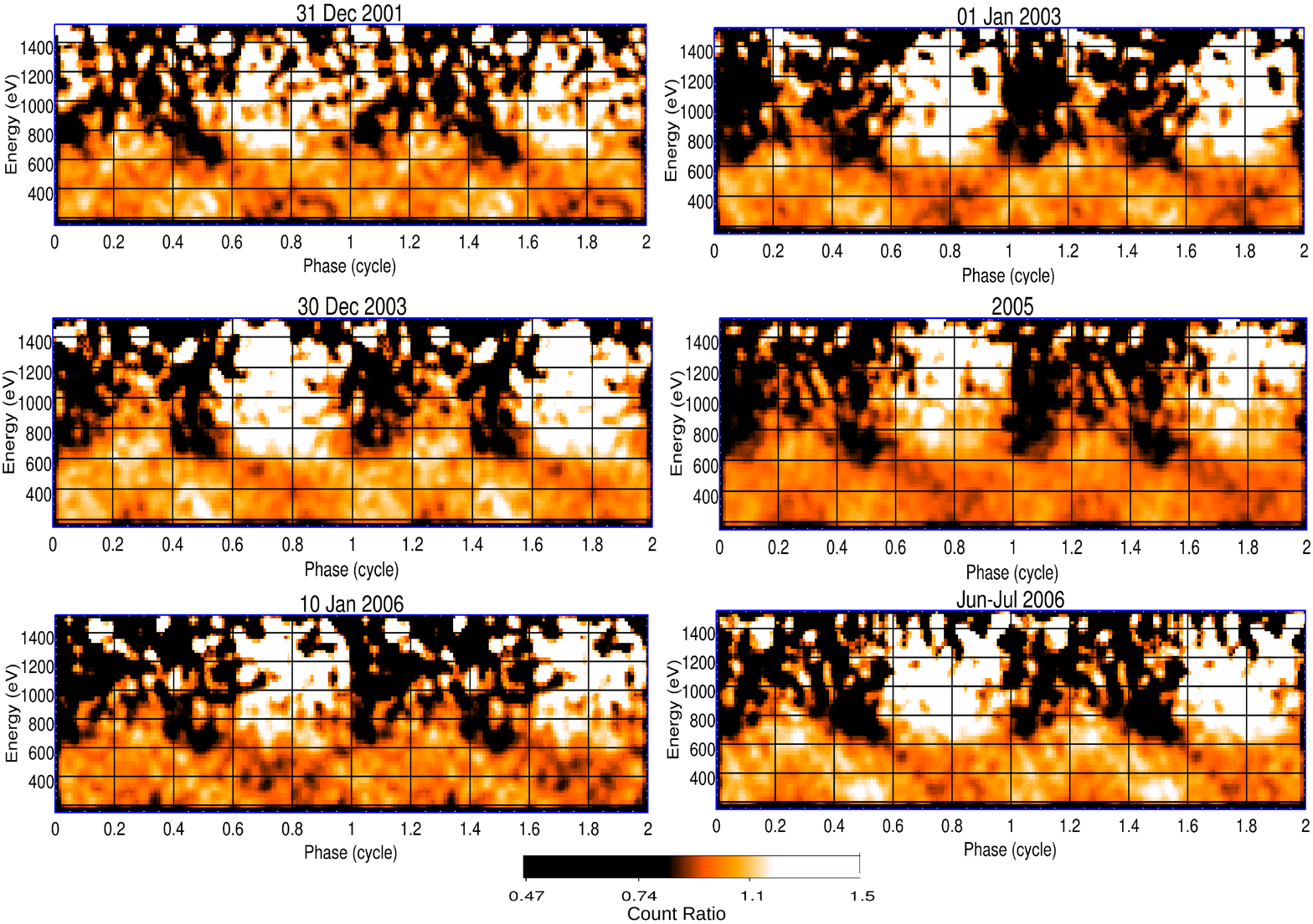}
\end{center}
\caption{Normalized energy versus phase images of \xdin\, obtained by binning the EPIC-pn source counts into 100 phase bins and 25-eV-wide energy channels for the observations considered in the spectral analysis.}
\label{fig:fiamme}
\vskip -0.1truecm
\end{figure*}

\begin{figure}
\begin{center}
\includegraphics[width=8.5cm]{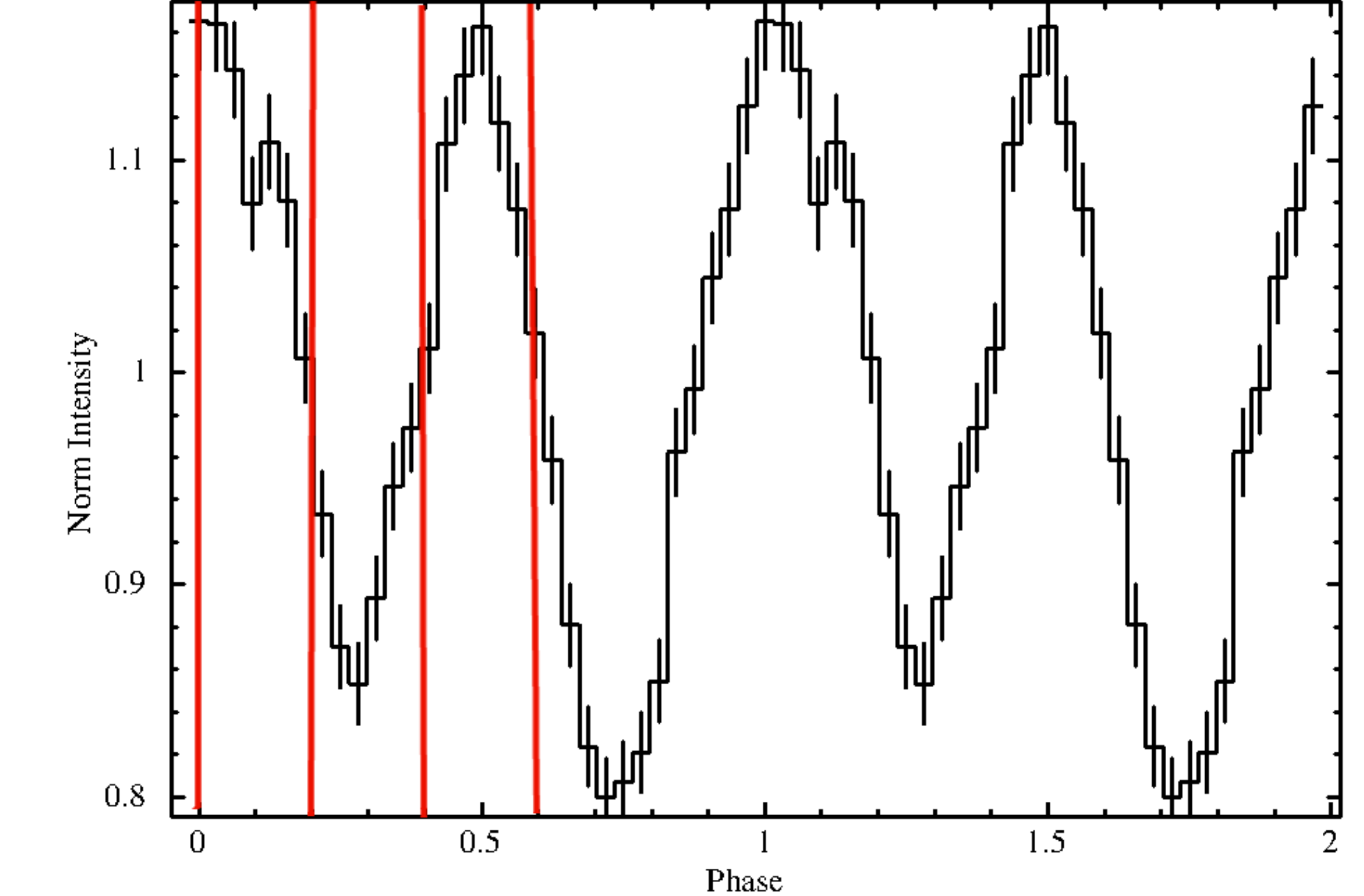}
\end{center}
\caption{Pulse profile of \xdin\, obtained from the longest observation. The vertical red lines highlight the phase intervals 0--0.2 and 0.4--0.6.  }
\label{fig:pulse_profile}
\vskip -0.1truecm
\end{figure}

To examine the spectral variations as a function of the star rotational phase, normalized energy versus phase images were created by binning the EPIC-pn source counts into 100 rotational phase bins and 25-eV-wide energy channels, and then normalizing to the phase-averaged energy spectrum and pulse profile. This method allows us to look for narrow spectral features that vary along the phase, such as the phase-dependent absorption feature detected in the magnetar SGR\,0418+5726 \citep{2013Natur.500..312T}, without making assumptions about the spectral energy distribution. The images (Figure \ref{fig:fiamme}) show two features in the phase intervals 0--0.2 and 0.4--0.6 (see the pulse profile in Figure \ref{fig:pulse_profile}), which are produced by a lack of counts with respect to the nearby energy channels and are evident at energies higher than $\sim$ 0.6\,keV. These might correspond to phase-variable absorption features that are present only in a limited interval of the rotational cycle.  


Based on the hint given by the normalized images, we performed a pulse-phase spectroscopy by dividing the phase cycle of each of the six available spectra into five equal bins, each of width 0.2 in phase. For each phase bin we fitted the spectra jointly in the 0.2--1.2\,keV energy band, fixing the column density at the phase-averaged value and constraining the parameters to be the same across all the observations, apart from relative normalization between different modes. As for the phase-averaged spectral analysis we assigned a 1.5\% systematic error term to each spectral channel.

\begin{figure}
\begin{center}
\includegraphics[width=8.5cm]{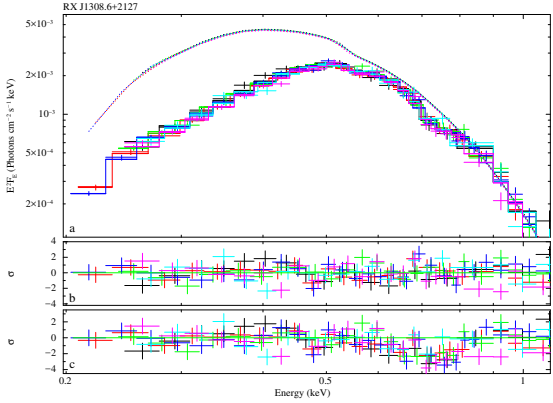}
\end{center}
\caption {Panel (a): unfolded phase-resolved spectra of the 0.4-0.6 phase interval fitted simultaneously. The solid lines represent the best-fitting model to the data (an absorbed blackbody plus two absorption Gaussian profiles), while the dashed lines represent only the blackbody component. The width of the phase-dependent line is free to vary. Panel (b): residuals with respect to this model. Panel (c): residuals after setting the normalization of the phase-dependent line to zero. The data were rebinned for plotting purposes only.}
\label{fig:04-06}
\vskip -0.1truecm
\end{figure}

We started the fitting procedure by considering the spectra relative to the phases 0--0.2 and 0.4--0.6. Both spectra were first modelled with an absorbed blackbody with a Gaussian in absorptioin ({\tt phabs*(bbodyrad+gauss)} in XSPEC, which we dub as the `continuum model' in the following), in order to properly model the broad feature (see Figure \ref{fig:04-06}). The broad feature parameters were not fixed at the values obtained from the phase-averaged spectral analysis because a dependence of the feature energy centroid on the rotational phase is reported \citep{2011A&A...534A..74H}. The best-fit with the above-mentioned model gives a \rchisq\, of 1.12 for 141 dof and 1.20 for 139 dof for the spectra in the 0--0.2 and 0.4--0.6 phase intervals, respectively. As shown in Figure \ref{fig:04-06}, panel c, the residuals in the spectrum relative to phase bin 0.4--0.6 show a discrepancy between the model and the data around $\sim$ 0.7\,keV, which is less prominent for the 0--0.2 phase-resolved spectrum. Therefore, as a next step, we introduced in the model an absorption component in the form of the additive {\tt gauss} model. This leads to an improvement in the shape of the residuals and in the \rchisq\, especially for the spectrum at phase 0.4--0.6 (see Figure \ref{fig:04-06}, panel b), while for the spectrum corresponding to the phase interval 0--0.2 the inclusion of an absorption line improves the fit, although not by a statistically significant amount. For both datasets, the parameters of the continuum model are consistent with the previous results; the best-fit values for the second Gaussian feature are the following: line energy E$_2$ = 778$^{+32}_{-27}$\,eV and equivalent width eqw$_2$ = 11$_{-5}^{+6}$\,eV in the 0--0.2 phase range, E$_2$ = 736$^{+18}_{-16}$\,eV and eqw$_2$ = 15$\pm$5\,eV in the 0.4--0.6 interval. The line width was fixed at 0\,eV. Formally this means that it is compatible with the spectral energy resolution of the pn camera at $\sim$ 700\,eV, which is about 100\,eV for single pixel events. If the width is free to vary, an upper limit of 86\,eV is obtained, quoted at 90\% confidence level. Table \ref{tab:all_pps} summarizes the results of the phase-resolved spectroscopy for the simultaneous fit. We do not report the parameters for the phase-variable narrow feature when its inclusion has a significance $< 2\sigma$, estimated applying the $F$-test (see, however, Section \ref{sect:MC}).

As shown in Table \ref{tab:all_pps}, the energy of the broad feature (E$_{\rm 1}$ $\sim$ 110--260\,eV) and the blackbody temperature (kT$_{\rm BB}$ $\sim$ 75--82\,eV) depend on the rotational phase. 

We investigated the significance of the narrow feature by modelling the continuum spectrum under different assumptions on the broad line shape ({\tt gabs} model) and using different values for the abundances and cross-sections of the photoelectric absorption model (i.e \citealt{1996ApJ...465..487V} and \citealt*{2000ApJ...542..914W}). We found that the properties of the phase-variable feature are not influenced by the choice of the cross sections, abundances and different spectral model for the broad line.


\begin{table*}
\begin{center}
\caption{Best-fit spectral parameters for a simultaneous fit of the phase-resolved spectra.}
\begin{tabular}{lccccc}
\hline \vspace{0.2cm}
Parameter$^a$	    & 0--0.2 	&	0.2--0.4 & 0.4--0.6 & 0.6--0.8 & 0.8--1 \\							
\hline 
\multicolumn{6}{c}{BB+GAUSS} \\
\hline
    
  kT$_{\rm BB}$ (eV) & 77.7$^{+1.8}_{-2.0}$ & 75.4$^{+2.2}_{-2.5}$ & 84.9$^{+1.3}_{-1.4}$ & 75.6$^{+2.1}_{-2.7}$ & 84.9$^{+1.8}_{-2.0}$   \\

  $R_{\rm BB}$ (km) & 4.3$\pm$0.5 & 5.6$\pm$1.0 & 2.6$\pm$0.1 & 5.8$\pm$1.1 & 3.4$\pm$0.3  \\
   
Flux$^b$  & 3.34$^{+0.04}_{-0.09}$  & 3.67$^{+0.15}_{-0.09}$ & 3.10$^{+0.05}_{-0.07}$ & 3.68$^{+0.07}_{-0.06}$ & 3.69$\pm$0.06  \\

Unabs. Flux$^b$  & 7.42$\pm$1.10 & 7.69$^{+1.70}_{-1.01}$ & 6.63$^{+0.63}_{-0.36}$ & 8.26$^{+1.35}_{-1.39}$ & 7.77$^{+0.55}_{-0.76}$  \\

E$_{\rm 1}$ (eV) & 173$^{+32}_{-39}$  & 107$^{+44}_{-54}$ & 256$^{+22}_{-28}$ & 109$^{+41}_{-59}$ & 198$^{+30}_{-36}$ \\

$\sigma_{\rm 1}$ (eV) & 143$^{+13}_{-12}$ & 169$^{+15}_{-14}$ & 105$^{+13}_{-11}$ & 168$^{+16}_{-13}$ & 146$^{+14}_{-12}$  \\

Eq Width$_{1}$ (eV) & 182$^{+2}_{-8}$ & 204$^{+2}_{-35}$ & 128$^{+10}_{-14}$ & 203$^{+2}_{-5}$ & 171$^{+11}_{-29}$ \\


NHP$^{c}$  & 1.6$\times 10^{-1}$ & 1.5$\times 10^{-1}$ & 5.4$\times 10^{-2}$  & 1.3$\times 10^{-1}$ & 2.8$\times 10^{-3}$ \\
  
$\chi^2_\nu$ & 1.12 & 1.12 & 1.20 & 1.13 & 1.35\\

dof & 141 & 149 & 139 & 147 & 150 \\
  \hline
  \hline
  \multicolumn{6}{c}{BB+GAUSS+GAUSS$^d$} \\
  \hline
     
    
   kT$_{\rm BB}$ (eV) & 77.2$^{+2.2}_{-2.3}$ &  & 84.3$^{+1.7}_{-2.1}$ &  & \\
     
 $R_{\rm BB}$ (km) & 4.6$\pm$0.8 &  & 2.8$\pm$0.3 &  &    \\
  
  E$_{\rm 1}$ (eV) & 144$^{+45}_{-52}$  &  & 224$^{+35}_{-52}$ &  &  \\

$\sigma_{\rm 1}$ (eV) & 155$^{+19}_{-16}$ & & 123$^{+21}_{-17}$ & & \\

Eq Width$_{1}$ (eV) & 189$^{+4}_{-5}$ &  & 135$^{+16}_{-3}$ &   &   \\

  E$_{\rm 2}$ (eV) & 778$^{+32}_{-27}$  & & 736$^{+18}_{-16}$ &   &   \\


  
  Eq Width$_{\rm 2}$ (eV) & 11$^{+6}_{-5}$ &  & 15$\pm$5 &  &   \\


  NHP$^{c}$  &  3.0$\times 10^{-1}$ &   & 3.4$\times 10^{-1}$ &  &\\

   $\chi^2_\nu$ & 1.06 &  & 1.05 & &   \\

dof & 139 & & 137 &  & \\
\hline \hline
\end{tabular}
\begin{list}{}{}
\item[$^{a}$] The N$_{\rm H}$ was frozen at the value obtained for the phase averaged spectra: N$_{\rm H}$ = 3.0$\times 10^{20}$\cm2 . 
\item[$^{b}$] Fluxes are calculated in the 0.1-10 keV energy range and in units of 10$^{-12}$ \ergscm2 with the model {\tt cflux} .
\item[$^{c}$] NHP is the Null Hypothesis Probability.
\item[$^{d}$]The width for the phase-dependent feature is consistent with the energy resolution of the pn camera.
\end{list}
\label{tab:all_pps}
\end{center}
\end{table*}

\section{\xdin: Monte Carlo simulations and significance of the narrow feature}
\label{sect:MC}

In order to have a preliminary estimate of the significance of the phase-variable absorption line, we applied the $F$-test. Taking into account the five trials corresponding to the number of fitted phase-resolved spectra the $F$-test probability is 4$\times$10$^{-2}$ ($\sim$ 2$\sigma$ confidence level) and 14$\times$10$^{-5}$ ($\sim$ 4$\sigma$ confidence level) for the spectra in the phase bin 0--0.2 and 0.4--0.6, respectively. Even though the $F$-test is widely used to estimate the line significance, it was shown how a more rigours approach requires the use of Monte Carlo (MC hereafter) simulations \citep{2002ApJ...571..545P}, which we report below.

To verify the presence of this phase-dependent absorption feature in the phase interval 0.4--0.6 we ran MC simulations, as suggested by \citet{2002ApJ...571..545P}. We simulated 10$^5$ spectra based on the best-fit {\tt phabs*(bbodyrad+gauss)} model (the null model), by means of the XSPEC {\tt fakeit} command. We fixed the hydrogen column density and the parameters of the broad Gaussian profile (energy, width and normalization) at their fitted values, while the blackbody temperature and normalization were drawn randomly from a Gaussian distribution centred on the best-fit with a width equal to the derived error, using the {\tt simpars} command. We assigned the response and background files of the phase-resolved 0.4--0.6 spectrum relative to the longest data set (obtained by merging all the observations performed during 2005) to the simulated spectra. The exposure time was set equal to the sum of the exposure times of each phase-resolved spectrum in the 0.4--0.6 phase bin (about 26\,ks). Each simulated spectrum was then binned using the same grouping algorithm as the real data, at least 30 counts per bin and with an oversampling factor of 5. 

\begin{figure*}
\begin{center}
\includegraphics[width=8.5cm]{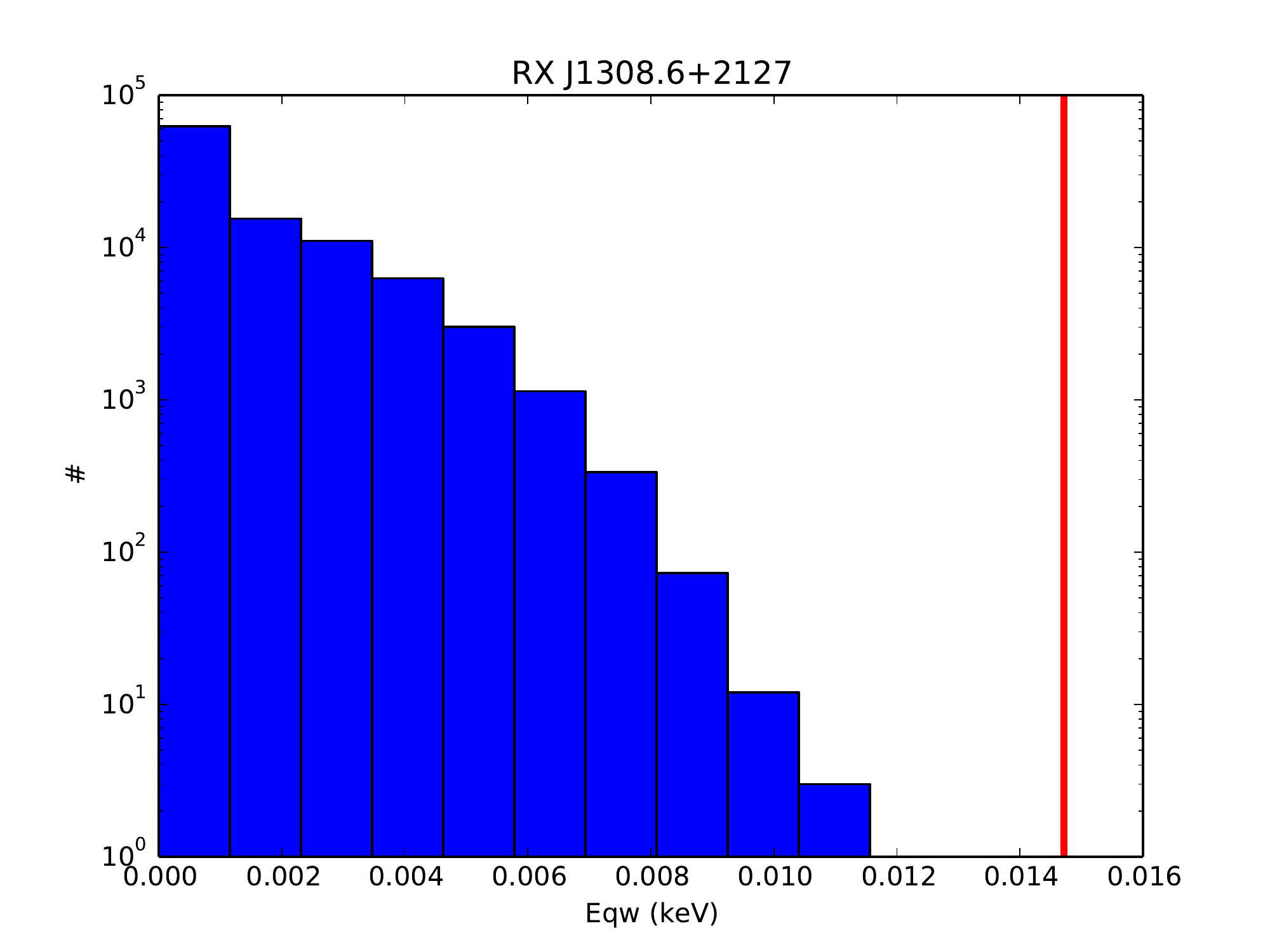}
\includegraphics[width=8.5cm]{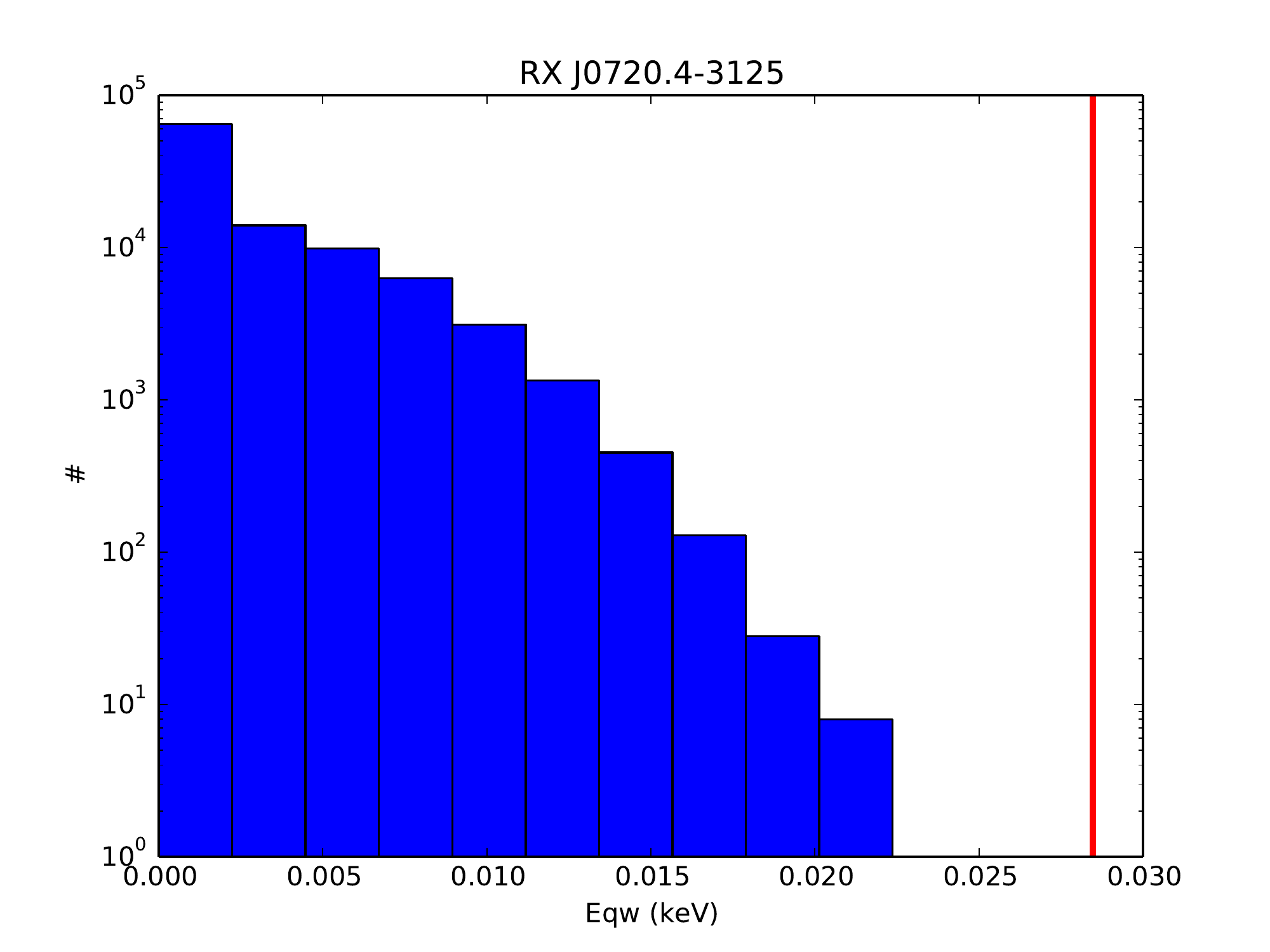}
\end{center}
\caption{Monte Carlo simulations of 10$^5$ spectra. The solid red line represents the observed value for the equivalent width. Left: result for \xdin. Right: result for RX J0720.4-3125. }
\label{fig:MC_sim}
\vskip -0.1truecm
\end{figure*}

We then fitted the simulated datasets with an alternative model, {\tt phabs*(bbodyrad+gauss+gauss)}. Given that we are testing for the presence of the 737\,eV absorption line in each of the `fake' spectra, we froze the energy of the narrow Gaussian line at this value, leaving all other parameters free to vary in the fit. In most of the simulated spectra the best result we could get was an upper limit for the equivalent width of the narrow spectral feature. The results of the simulation are shown in Figure \ref{fig:MC_sim}, left panel: none of the simulated spectra has an equivalent width greater than that obtained from the real data (red line; see also Table \ref{tab:all_pps}). The probability of the feature being a fluctuation is $<$ 10$^{-5}$, which can be interpreted as a $p$-value. We note that the results of the simulation rely on the assumption that the null model holds, therefore a $p$-value of 10$^{-5}$ means that the feature is unlikely to have occurred by chance if the null model is true. A $p$-value of 10$^{-5}$ corresponds to a 4.6$\sigma$ confidence level, confirming the detection of the phase-variable absorption feature at 737\,eV in the X-ray spectrum of \xdin.\\

For completeness, we repeated the same procedure for the phase-dependent line that we discovered in the spectrum of the XDINS RX\,J0720.4-3125 \citep{2015ApJ...807L..20B}. We again simulated 10$^5$ spectra under the null model hypothesis ({\tt tbabs*bbodyrad} model) reported in \citet[Table\,2]{2015ApJ...807L..20B}, and with the same exposure time ($\sim$ 10\,ks) as the phase-resolved spectrum where we detected first the line, corresponding to the observation performed on 2003 May 2. As shown in Figure \ref{fig:MC_sim}, right panel, all the simulated datasets have an equivalent width lower than the observed value of 28\,eV (red line in the plot), therefore this result further confirms the absorption line presence at 745\,eV\footnote{ During the referee process a paper by \citet{2017arXiv170207635H} claiming a new period for RX\,J0720.4-3125 was published. The claimed period is twice of the one reported in literature, which we used in our work \citep{2015ApJ...807L..20B}. Our result is not affected by this new measurements. Folding the light curve at the new most plausible period, the feature appears in two phase intervals, as expected.}.

\section{Search for similar lines in other XDINS\lowercase{s}}
\label{sect:UL}

\begin{figure}
\begin{center}
\includegraphics[width=8.5cm]{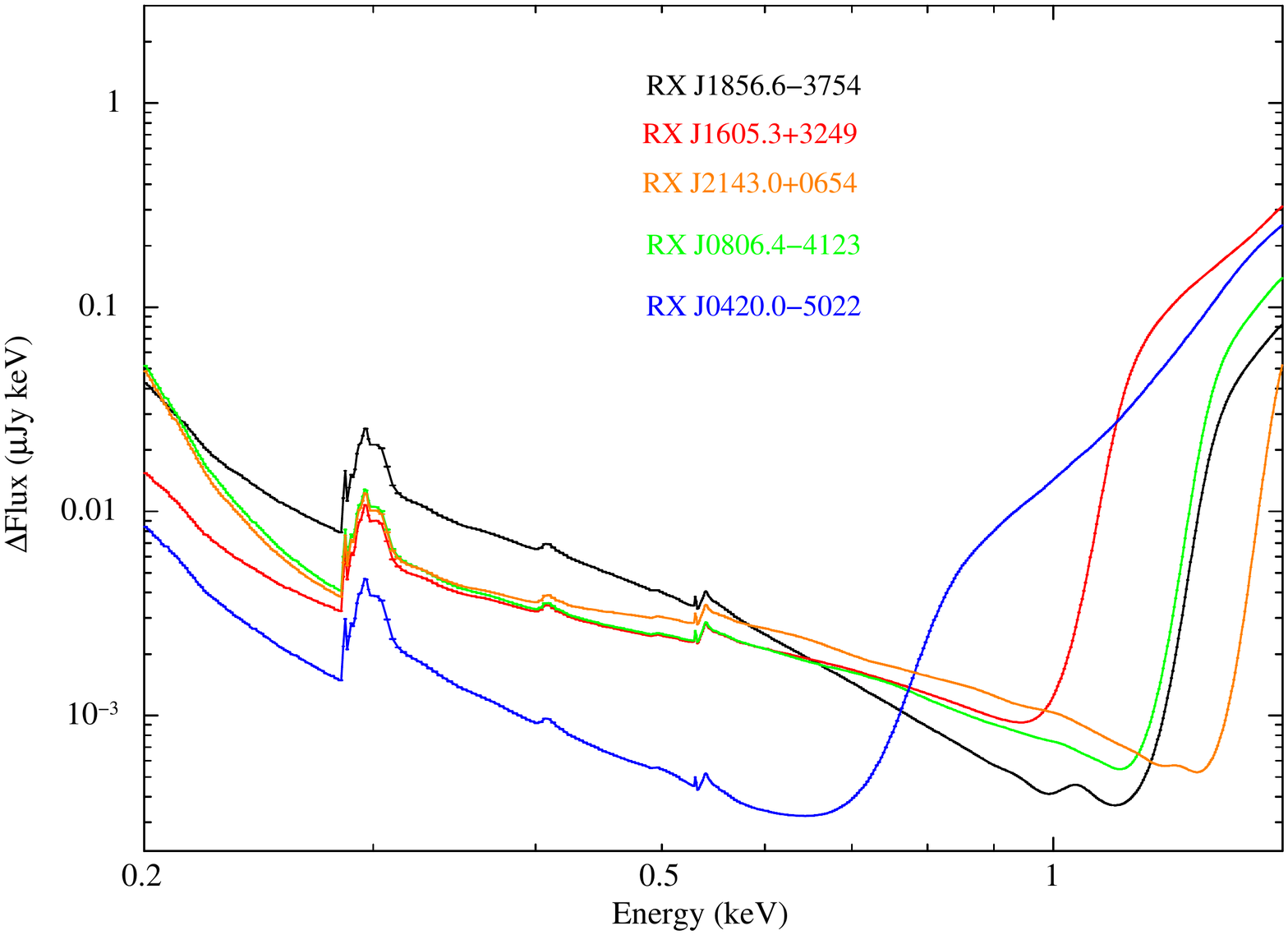}
\end{center}
\caption{ 1$\sigma$ upper limit on a detectable flux variation with respect to the continuum model as a function of energy.}
\label{fig:upperlimit}
\vskip -0.1truecm
\end{figure}

We proceeded with a systematic investigation for similar phase-dependent features using the longest \xmm\, observations of all the other XDINSs, focusing on data taken with the EPIC-pn camera because of the larger counting statistics. The data were reduced consistently with the criteria reported in Section \ref{subsect:reduction}. A log of the studied sources with the corresponding analysed observations is given in Table \ref{tab:sources}. We created the energy versus phase images by binning the pn source counts into 50 rotation phase bins and 20-eV-wide energy channels. No particular features from the normalized images have been observed in these sources. For the phase-averaged spectra we derived 3$\sigma$ upper limits on the equivalent width of a Gaussian line in absorption with width $\sigma_{line}$ = 0 (narrower than the pn camera spectral energy resolution) and $\sigma_{line}$ = 100 eV. The results are reported in Table \ref{tab:upperlimit}. A summary is shown in Figure \ref{fig:upperlimit}, where we plot the 1$\sigma$ upper limit on a detectable flux variation with respect to the continuum model as a function of energy, which can be related to the detectability of the spectral lines as broad as the energy-dependent instrument resolution on top of the continuum flux. 

\begin{table*}
\begin{center}
\caption{Summary of the sources and the corresponding observations where we searched for phase-dependent spectral features.}
\label{tab:sources}
\footnotesize{
\begin{tabular}{@{}lclcccc}
\hline
\hline
Source & Obs. ID  	&  Obs. Date 	& Read-out mode  	& Duration & Count rate$^{a}$        & Unabs. Flux$^b$\\ 
	   &           	& YYYY-MM-DD 	& 					& (ks)     & (counts s$^{-1}$) & ( \ergscm2 ) \\
\hline
RX J1856.5-3754 & 0412601501 & 2011 Oct 05 & SW & 118.1 & 6.78(2) & 2.00 $\times$ 10$^{-11}$ \\
RX J1605.3+3249 & 0671620101 & 2012 Mar 06 & FF & 60.4  & 4.33(1) & 1.20 $\times$ 10$^{-11}$ \\
RX J2143.0+0654 & 0201150101 & 2004 May 31 & SW & 30.4  & 2.20(1) & 4.44 $\times$ 10$^{-12}$ \\
RX J0806.4-4123 & 0141750501 & 2003 Apr 24 & FF & 33.6  & 1.72(1) & 3.89 $\times$ 10$^{-12}$ \\
RX J0420.0-5022 & 0141751101 & 2003 Jan 19 & FF & 22.4  & 1.172(1) & 1.46 $\times$ 10$^{-12}$  \\
\hline
\hline
\end{tabular}}
\begin{list}{}{}
\item[$^a$] Count rates refer to the spectra extracted with a circle region with PATTERN $\leq$ 4 in the 0.1--2 keV energy band; errors are quoted at the 1$\sigma$ confidence level.
\item[$^b$] Unabsorbed fluxes are calculated in the energy range 0.1--2 keV. 
\end{list}
\end{center}
\end{table*}

\begin{table}
\begin{center}
\caption{3$\sigma$ upper limits on the equivalent width of a Gaussian feature in absorption with a width $\sigma_{line}$ = 0 (narrower than the pn spectral energy resolution) and 100 eV, derived for the observations listed in Table \ref{tab:sources}. The limits are calculated in the 0.1--1.5 keV energy range for all the sources. }
\label{tab:upperlimit}
\footnotesize{
\begin{tabular}{@{}lcc}
\hline
\hline
Source & $\sigma_{line}$ = 0 (eV) & $\sigma_{line}$ = 100 (eV)   \\ 

\hline
RX J1856.5-3754 & $<$10 & $<$14 \\
RX J1605.3+3249 & $<$18 & $<$23 \\
RX J2143.0+0654 & $<$21 & $<$28 \\
RX J0806.4-4123 & $<$32 & $<$69 \\
RX J0420.0-5022 & $<$54 & $<$44 \\
\hline
\hline
\end{tabular}}

\end{center}
\end{table}

\section{Discussion}
\label{sect:disc}






Thanks to a careful re-analysis of all the available \xmm\, observations of XDINSs, we found a phase-dependent absorption feature in the X-ray spectrum of RX J0720.4-3125 \citep{2015ApJ...807L..20B}. Here we report the discovery of another such feature in \xdin\,, and present upper limits for such narrow lines for the other XDINSs. The new feature was first detected visually, by inspecting normalized energy versus phase images in two different phase intervals. A detailed phase-resolved spectral analysis, and Monte Carlo simulations, confirmed its presence at $>4.6\sigma$ confidence level in the phase range 0.4-0.6. The phase-resolved spectra corresponding to the observations between December 2001 -- July 2006 were fitted simultaneously, assuming the feature being stable over the period covered by the datasets. 
Our searches for the feature in \xmm\, RGS and \cxo\, data were inconclusive owing to the much poorer counting statistics compared to that available from the EPIC-pn camera.

For both RX J0720.4-3125 and \xdin\, the feature is detected only in a narrow phase interval ($\sim$ 20\% of the pulsar rotation), the line energy is $\sim$ 750 eV, and the width is consistent with the spectral energy resolution of the pn camera, about 100 eV for single pixel events at the centroid energy. Interestingly a similar phase-variable absorption feature (although with a larger energy shift with phase) has been detected in two low-field magnetars, SGR 0418+5729 and SWIFT J1822.3-1606 \citep{2013Natur.500..312T, 2016MNRAS.456.4145R}. In these sources the line energy is higher ($\geq$ 2 keV) and shows a strong variation in phase, which is not evident in both XDINSs possibly because the source counts become background dominated at energies just above that of the feature ($\sim$ 1 keV). These findings strengthen the evolutionary link between magnetars and the seven thermally-emitting isolated neutron stars, thought to be aged magnetars according to the most updated magneto-thermal evolutionary models \citep{2013MNRAS.434..123V}.

In the light of the similarities with SGR 0418+5729 and RX J0720.4-3125, the feature might be explained invoking proton cyclotron absorption/scattering \citep{2013Natur.500..312T}. This interpretation quite naturally accounts for the sharp variation with phase if there are small-scale ($\approx 100\ \mathrm m$) magnetic structures close to the neutron star surface. The proton cyclotron energy measured by a distant observer is $E_{\rm c} = (e B \hslash / m_{\rm p})/(1+z) = 0.63 B_{14} / (1+z)$ keV, where $B_{14}=B/10^{14}\ {\rm G}$, $z = 2 G M_{\rm NS} / R_{\rm NS} c^2 \sim 0.4$ (assuming a neutron star mass of $M_{\rm NS} =  1.4 M_{\odot}$ and a radius of $R_{\rm NS} =10$\,km) and $1+z$ is the gravitational redshift. The implied magnetic field in the loop is then $B_{\rm loop} \sim 1.7 \times 10^{14}$\,G, about a factor 5 higher than the dipole field at the equator, $B_{\rm dip}\sim 3.4\times10^{13}$\,G \citep{2005ApJ...635L..65K}. If confirmed, this interpretation supports the scenario according to which the magnetic field of highly magnetized neutron stars is complex with deviations from a pure dipole on small scale, such as localized high $B$-field bundles. Unfortunately, an estimate of the exact timescale of the evolution of these small magnetic structures is beyond the capabilities of current simulations, and it is highly dependent on many parameters such as the B-field geometry at birth, the crustal conductivity, the origin of the helicity of the specific structure, etc. \citep{1993ApJ...408..194T,1995MNRAS.275..255T,2009ApJ...703.1044B}. In this scenario, these uncertainties do not allow us to make any prediction on the long-term stability of this absorption feature.


The absorption feature we report in this work is likely unrelated to the broad one detected at $\sim$ 270\,eV in the phase-averaged spectrum, which,  as in the case of RX J0720.4-3125, shows a phase dependence too \citep{2011A&A...534A..74H,2006A&A...451L..17H}. If we assume that both spectral features are produced by proton cyclotron resonance absorption/scattering, the latter yields a lower magnetic field, comparable with the spin-down value of the dipole ($\sim 6 \times 10^{13}$\,G), and might be linked to the large-scale field component. 

Except for RX J1856.5-3754 (and possibly RX J0420.0-5022), all XDINS exhibit broad absorption features in their X-ray spectra, but the strong phase variability detected in the new narrow features of \xdin\, and RX J0720.4-3125 sets them apart. The origin of absorption features in XDINS spectra is still unclear and bound-bound/bound-free transitions in strongly magnetized atoms/molecules have been invoked as an alternative explanation to resonant absorption/scattering \cite[see e.g.][]{2009ASSL..357..141T}. However, this interpretation is not without problems. Absorption from atoms in a strong magnetic field implies that an atmosphere needs to be present around these stars. This is in many cases not supported by observations, since atmospheric spectral models provide an unsatisfactory description of XDINS X-ray data. Also, the line energy in RX J2143.0+0654 \cite[RBS 1774;][]{2005ApJ...627..397Z} is too high to be explained by transitions in light element atoms. In the case of the narrow, phase-variable lines, the reprocessing must occur in a very limited region in order to account for the strong dependence on rotational phase. Either primary emission is also coming from the same small region, or it is the ``atmosphere'' which covers only a small part of the surface, while emission comes from a much larger region. The first scenario is not supported by observations because of the broad pulse profile and the quite large radiation radius. The second might work but it is rather difficult to understand how some absorbing material is confined only on top of a small surface area. Besides, both in \xdin\ and  RX J0720.4-3125  the line energy is too high to be explained by atomic transitions in light elements atoms.\\

A search for similar phase-dependent features has been performed through
visual inspection of the normalized phase-energy images for some high-B
pulsars and other magnetars, but without conclusive results so far. At
variance with the XDINSs, the spectra of high-B pulsars and young
magnetars are dominated by magnetospheric non-thermal emission, either due
to synchrotron radiation or resonant cyclotron scattering. Therefore a
spectral line due to scattering near the stellar surface can be easily washed
out. Furthermore, in young magnetars the difference in strength between the
large scale dipole field and the field concentrated in small structures close
to the surface is not expected to be large, hence the formation of narrow
lines might be inhibited. On the other hand, in older
systems, such as the XDINSs and the low-field magnetars, small scale
magnetic structures might have field strengths that can be a factor
10--100 stronger than the (decayed) dipolar component. Together with
their typical soft thermal emission, this might increase the chances
of the formation and detection of narrow resonant cyclotron features.


\section{Conclusions}
\label{sect:concl}

A detailed phase-resolved spectral analysis of archival \xmm\, EPIC-pn data allowed us to detect a narrow phase-variable absorption feature in the X-ray spectrum of \xdin\, and to derive upper limits for the other XDINSs, in which such a study was not carried out yet (a feature with similar properties was previously reported in RX J0720.4-3125 \citep{2015ApJ...807L..20B}). In the case of RX J1308.6+2127 the spectral feature is present only in 20\% of the phase cycle with an energy of $\sim$ 740\,eV and an equivalent width of $\sim$ 15\,eV. 


The characteristics of the features point towards the proton cyclotron absorption/scattering interpretation, which explains the strong dependence on the pulsar rotation if localized magnetic bundles are present close to the stellar surface. This view provides evidence for deviations from a pure dipole magnetic field on small scales for highly magnetized neutron stars.

The proposed mission {\em Athena}, currently scheduled for launch in 2028, will play a key role for the characterization of phase-dependent absorption features thanks to its high collective area (about 2 m$^2$ at an energy of 1 keV; \citealt{2013sf2a.conf..447B}).
The combination of its large effective area and high spectral resolution will enable to study these phase-variable features in more detail, and possibly detect even the most subtle lines in other highly magnetized neutron stars.

\section*{Acknowledgements}
AB, NR and FCZ are supported by an NWO Vidi Grant (PI: Rea), and by the European COST Action MP1304 (NewCOMPSTAR). NR is also supported by grants AYA2015-71042-P and SGR 2014-1073.

\bibliographystyle{mnras}
\bibliography{reference}

\end{document}